\documentclass[prd, preprint, nofootinbib]{revtex4}
\usepackage[dvipdfmx]{graphicx}
\usepackage{amsfonts}
\usepackage{latexsym}
\usepackage{amssymb}
\usepackage{amsmath}
\usepackage{epsfig}
\usepackage{comment}
\usepackage{here}

\usepackage{color}
\usepackage{caption}
\newcommand\ba{\begin{eqnarray}}
\newcommand\ea{\end{eqnarray}}

\begin{document}

\title{Small field axion inflation with sub-Planckian decay constant}

\author{$^1$Kenji Kadota, $^2$Tatsuo Kobayashi, $^3$Akane Oikawa, $^2$Naoya Omoto, $^3$Hajime Otsuka, and  $^2$Takuya H. Tatsuishi}
 \affiliation{ $^1$Center for Theoretical Physics of the Universe, Institute for Basic Science, Daejeon 305-811, Korea\\
$^2$Department of Physics, Hokkaido University, Sapporo 060-0810, Japan\\
$^3$Department of Physics, Waseda University, Tokyo 169-8555, Japan}



\begin{abstract}
We study an axion inflation model recently proposed within the framework of type IIB superstring theory, where we pay a particular attention to a sub-Planckian axion decay constant. 
Our axion potential can lead to the small field inflation with a small tensor-to-scalar ratio, and a typical reheating temperature can be as low as GeV. 
%
\end{abstract}

\pacs{}
\preprint{CTPU-16-15}
\preprint{EPHOU-16-008} 
\preprint{WU-HEP-16-12}

\vspace*{3cm}
\maketitle



\section{Introduction}

The moduli generically appear in superstring theory with compactification and 
their vacuum expectation values correspond to the size and shape of the compact space. The moduli fields hence can offer characteristic features in superstring theory on the six-dimensional compact space, and they can play important roles in particle phenomenology and cosmology of the four-dimensional low-energy effective field theory.

The moduli have perturbatively flat potential and their imaginary parts, axions, posses the shift symmetries. The moduli, in particular axions, hence are good candidates for the inflaton field driving the cosmological inflation. A well-known example includes the natural inflation \cite{Freese:1990rb} where the non-perturbative effects break the shift symmetry into the discrete one and induce the non-flat potential for the axion.
A notable requirement for the successful natural inflation model is the super-Planckian axion decay constant, i.e. $f \sim 5 M_{\rm p}$ ($M_{\rm p}$ denotes the reduced Planck scale  $M_{\rm p}=2.4 \times 10^{18} \ {\rm GeV}$) while a typical decay constant in superstring theory is sub-Planckian, $f \lesssim M_{\rm p}$ \cite{Svrcek:2006yi}. The possibilities for realizing a super-Planckian decay constant hence have been explored such as the studies on the alignment mechanism \cite{Kim:2004rp} and 
the one-loop effects \cite{Abe:2014pwa,Abe:2014xja}. Another interesting axion inflation scenario in superstring theory is the axion monodromy inflation \cite{Silverstein:2008sg}.\footnote{
See also Refs.~\cite{Kobayashi:2014ooa,Higaki:2014sja}.}

These axion inflation models in the string theory discussed in the literature typically involve the super-Planckian inflaton amplitudes and a potentially large tensor-to-scalar ratio $r$ is featured for the large field excursion $\Delta \phi$ as \cite{Lyth bound}
\begin{eqnarray}
\frac{\Delta \phi}{M_{\rm p}} \simeq {\mathcal O}(1) \times \left(\frac{r}{0.01}\right)^{1/2}.
\label{eq:Lyth bound}
\end{eqnarray}

On the contrary, in this paper, we study the small field axion inflation where the field excursion of the axion inflaton 
is small compared with $M_{\rm p}$.\footnote{See, e.g. Refs.~\cite{Peloso:2015dsa,Kobayashi:2016mzg}.} 
The tensor-to-scalar ratio can be consequently small and, in our string axion models with a sub-Planckian axion decay constant, the reheating temperature can be as low as GeV.




For the illustrative purpose, we study in details the concrete axion inflation model which was recently derived within the framework of 
type IIB superstring theory \cite{Kobayashi:2015aaa}.
It is the extension of the work \cite{Hebecker:2015rya} to the compactification  with generic fluxes,
and the inflation potential consists of the mixture of polynomial functions and sinusoidal functions of the axion.\footnote{
See also Refs.~\cite{Parameswaran:2016qqq,Bizet:2016paj}.}.


The paper is organized as follows.
In section 2, we study the inflation dynamics for our axion inflation scenarios with a sub-Planckian axion decay constant and demonstrate that the axion inflation energy scale can be quite low compared to the conventional axion inflation scenarios with a super-Planckian axion decay constant.
In section 3 we study the reheating temperature in our model and
discuss the thermal history after the inflation, followed by the conclusion in Section 4.


\section{Axion inflation with a small axion decay constant}

We, in this section, present the axion inflation model based on type IIB superstring theory \cite{Kobayashi:2015aaa}.
In particular, we consider the inflation model with a sub-Planckian axion decay constant which can lead to a small tensor-to-scalar ratio $r$. We give the quantitative discussions for our axion inflation scenarios in terms of the slow-roll parameters  $\epsilon \equiv  \frac12 \left(\frac{V_{\phi}}{V} \right)^2$$, \eta\equiv \frac{V_{\phi \phi}}{V}$ in view of the Planck constraints \cite{Planck deta,plbi}
\begin{eqnarray}
  P_{\xi} &=& \left(\frac{H^2}{2\pi |\dot{\phi}|}\right)^2 = \frac{V}{24\pi^2 \epsilon}= 2.20 \pm 0.10 \times 10^{-9},
\label{eq:bounds on curvature perturbation}\\
n_s &= & 1+2\eta-6\epsilon= 0.9655 \pm 0.0062, \\
\label{eq:bounds on spectral-index}
r &= &16 \epsilon<0.12 .
\label{eq:tensor-to-scalar ratio}
\end{eqnarray}

\subsection{Axion inflation potential in type IIB string theory}
Recently, within the framework of type IIB superstring theory, the following form 
of axion potential was derived \cite{Kobayashi:2015aaa},
\begin{eqnarray}
V(\phi) = \Lambda_1\phi^2+\Lambda_2\phi \sin \left(\frac{\phi}{f}\right) +\Lambda_3\left(1-\cos \left( \frac{\phi}{f} \right) \right),
\label{eq:original axion potential}
\end{eqnarray}
where $\Lambda_{1,2,3}$ are constant, and $f$ is the axion decay constant.

We consider the flux compactification of type IIB superstring theory. We can, in general, stabilize all of the complex structure moduli and the dilaton by choosing proper 
3-form fluxes \cite{Giddings:2001yu,Gukov:1999ya}.
We here choose the 3-form fluxes such that only one of the complex structure moduli, $\Phi$, does not appear in the tree-level superpotential, 
while the other complex structure moduli as well as the dilaton are stabilized by the 3-form fluxes.\footnote{
We also assume that all of the K\"ahler moduli are stabilized by non-perturbative effects \cite{Kachru:2003aw} 
 and a proper uplifting scenario is available such as \cite{Dudas:2006gr}.}
However, the quantum corrections induce the superpotential, 
\begin{equation}
W = w_0 + (c + c'\Phi)e^{-\Phi/f},
\end{equation}
where $w_0,c, c'$ are constants determined by fluxes and vacuum expectation values of other moduli.
The Kahler potential of $\Phi$ also receives the correction,
\begin{equation}
\Delta K = \left(k+k'{\rm Re}(\Phi)\right) \cos ({\rm Im}(\Phi)/f)e^{-{\rm Re}(\Phi)/f},
\end{equation}
in addition to the tree-level K\"ahler potential $K= -\ln V$ with the volume $V$ of the internal manifold,
where $k$ and $k'$ are constants determined by fluxes and other moduli vacuum expectation values.
We assume that the real part of $\Phi$, ${\rm Re}(\Phi)$, is heavy, and integrating out ${\rm Re}(\Phi)$ leads to the above scalar potential  Eq. (\ref{eq:original axion potential}) for the axion $\phi = {\rm Im}(\Phi)$.\footnote{Recently, the authors of Ref.~\cite{Blumenhagen:2016bfp} pointed out that the light complex structure 
moduli appear in the explicit Calabi-Yau manifolds.}
Note that the superpotential as well as the K\"ahler potential includes 
the linear term, exponential term and their products. This is the origin of the mixture between polynomial functions and sinusoidal functions in the scalar potential.
See, for details, Ref.~\cite{Kobayashi:2015aaa}.
The natural scale for the decay constant would be of order $f \sim 1/2\pi$, even though one can expect a wider range depending on the vacuum expectation values of
the real parts of moduli corresponding to the sizes of cycles. For concreteness, in the following discussions, we mainly consider the range \footnote{Throughout this paper, we use the units where the reduced Planck scale $M_{\rm p}=2.4 \times 10^{18} \ {\rm GeV}=1$.}
\ba
 0.01 \le f \le 1.0. 
 \ea
 Note that we focus on a small axion decay constant which does not exceed the Planck scale, while a large axion decay constant has usually been explored in the previous literature on the axion inflation scenarios \cite{Kobayashi:2015aaa}.
The magnitudes and ratios of $\Lambda_{1,2,3}$ can vary depending on the flux magnitudes and vacuum expectation values of 
moduli \cite{Kobayashi:2015aaa},  and we here treat  $\Lambda_{1,2,3}$ as free parameters to make our discussions as general as possible.

\begin{figure}[htbp]
  \begin{center}
    \begin{tabular}{c}
                    
      \begin{minipage}{0.5\hsize}
        \begin{center}
                    \includegraphics[clip, width=6.5cm]{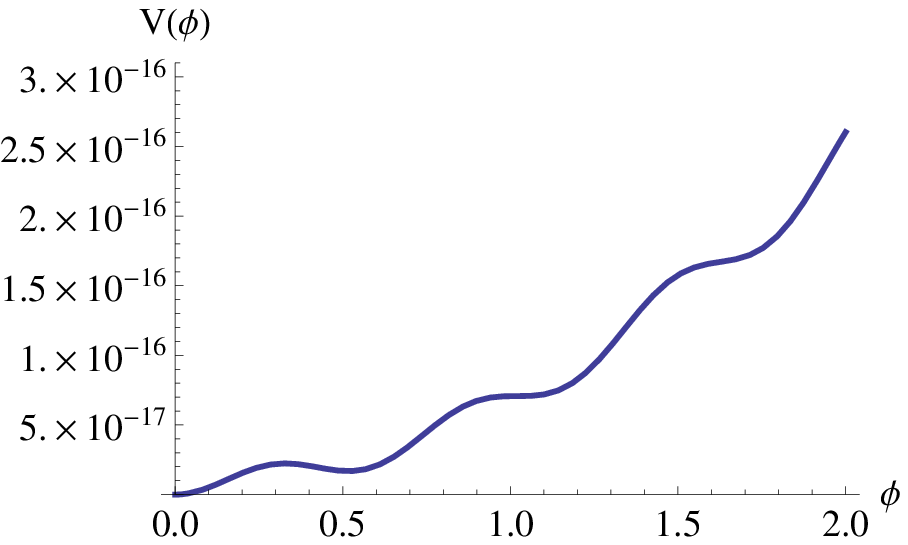}
                    \hspace{1.0cm}  $f=0.1$
        \end{center}
      \end{minipage}
              \begin{minipage}{0.5\hsize}
        \begin{center}
                    \includegraphics[clip, width=6.5cm]{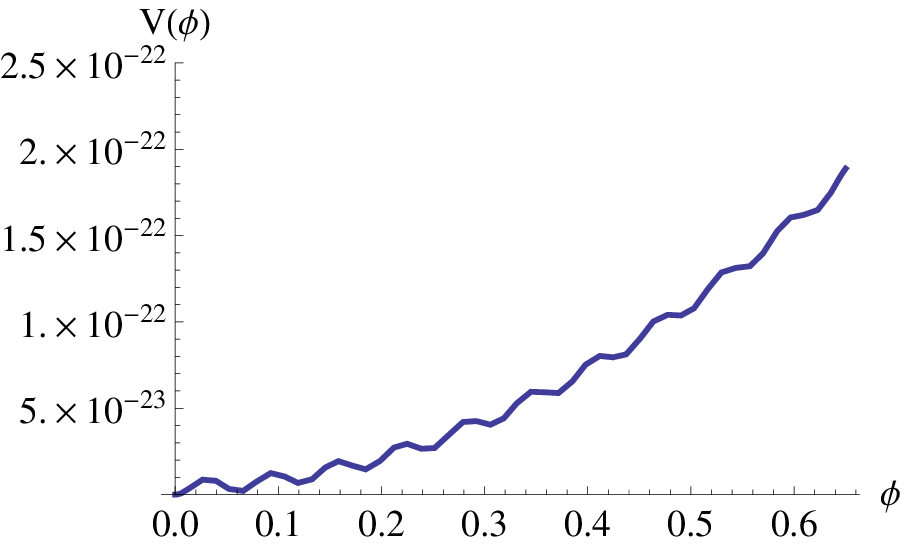}
             \hspace{1.0cm}  $f=0.01$
        \end{center}
      \end{minipage}
    \end{tabular}
    \caption{The axion inflation potentials with a sub-Planckian axion decay constant $f=0.1$(left) and $f=0.01$(right) 
    for the small field inflation (the field excursion $\Delta \phi < 1$ during the inflation).}
    \label{fig:Vsmallf}
  \end{center}
\end{figure}

This potential consists of a mixture of polynomial functions and sinusoidal functions. It reduces to the simple $\phi^2$ chaotic inflation when $\Lambda_2=\Lambda_3=0$, which is in a tight tension with the observations due to a large $r$ \cite{Planck deta}. 
For the non-vanishing $\Lambda_2$ and $\Lambda_3$, the potential consists of many bumps and plateaus, as shown in Fig. \ref{fig:Vsmallf}, as well as several local minima.
The form of the potential Eq.(\ref{eq:original axion potential}) heavily depends on the oscillation parameter $f$ which determines the width size of the flat plateau regime. A small $f$ leads to a high frequency potential with a small interval between each plateau, and our main focus is on a smaller value of $f$ making each flat plateau closer to each other.
The potential is shown in Fig. \ref{fig:Vsmallf} for $f=0.1$ and $f=0.01$, where, for concreteness, we chose $\Lambda_2/\Lambda_3=1, \Lambda_1/\Lambda_3=7.3$ for $f=0.1$ and $\Lambda_2/\Lambda_3=1, \Lambda_1/\Lambda_3=97$ for $f=0.01$. The inflation can occur on a flat plateau and we, in the following, study the inflation dynamics for our axion inflation scenarios with a sub-Planckian inflaton field excursion.

\subsection{Small field axion inflation}
The inflation can occur when an axion inflaton field slowly rolls over a flat plateau region in our axion potential. We shall demonstrate that the small field inflation can be realized for a small axion decay constant $f$ when an enough number of e-folds are induced for a sufficiently flat potential.
The first derivative of the potential is written by 
\begin{eqnarray}
V_{\phi}&=& \left( 2\Lambda_1+\frac{\Lambda_2}{f} \cos \left(\frac{\phi}{f} \right) \right)\phi +\left( \Lambda_2+ \frac{\Lambda_3}{f} \right) \sin \left( \frac{\phi}{f} \right),
\label{eq:first derivative of original potential} 
\end{eqnarray}
For our potential to become flat enough for a sufficient number of e-folds, we require $(V_\phi)^2 \ll V^2$, 
which is satisfied  for $\phi \sim 1$ and $f \ll 1$ (as well as $\cos(\phi/f), \ \sin (\phi/f) \sim {\cal O}(1)$) when
\begin{equation}
\Lambda_1 f \sim \Lambda_2  \sim \Lambda_3,
\label{eq:fine-tuning}
\end{equation}
with proper signs of $\cos(\phi/f)$ and $\sin (\phi/f)$.
Another condition $V_{\phi \phi} \ll V$ can also be satisfied in the same parameter region.
The consequent small inflaton field variation results in a small tensor-to-scalar ratio $r$ as estimated in the following.

For the inflaton variation $ \Delta \phi$ around $|V_\phi| \approx 0$ and $|V_{\phi \phi}| \approx 0$, the second derivative can be estimated as
\begin{equation}
V_{\phi \phi} \sim V_{\phi \phi \phi} \Delta \phi \sim \left(-\frac{\Lambda_3}{f^3} \sin\left( \frac{\phi}{f} \right) -  \frac{\Lambda_2}{f^3} \cos\left( \frac{\phi}{f} \right) 
\right) \Delta \phi.
\end{equation}
Note, for a small $f$, the terms with $f^{-3}$ can be dominant in the third derivative $V_{\phi \phi \phi}$.
For $V \sim \Lambda_1 \phi^2 \sim \Lambda_3/f$, with the relation (\ref{eq:fine-tuning}) and $\phi = {\cal O}(1)$,  
we estimate 
\begin{equation}
\eta \sim  \frac{\Delta \phi}{f^2}.  
\label{eq:eta-f}
\end{equation}
Demanding $\eta \ll 1$ results in $\Delta \phi \ll f^2$, which leads to $r \ll 0.01 \times f^4$ 
 from Eq. (\ref{eq:Lyth bound}).
Explicitly, we can write 
\begin{equation}
r \sim 10^{-6} \times f^4 \times \left( \frac{\eta}{0.01}\right)^2.
\label{eq:r-approx}
\end{equation}
In addition, we can estimate $\eta \sim 10^{-2}$ because $r = 16 \epsilon \ll 0.01$ and $2 \eta \approx n_s -1 \approx -0.03 $.
With this approximation, we estimate $r \sim 10^{-6} \times f^4$ and tensor-to-scalar ratio $r$ can be suppressed greatly as $f$ becomes small.

\begin{figure}[htbp]
  \begin{center}
    \begin{tabular}{c}
      \begin{minipage}{0.5\hsize}
        \begin{center}
          \includegraphics[clip, width=6.5cm]{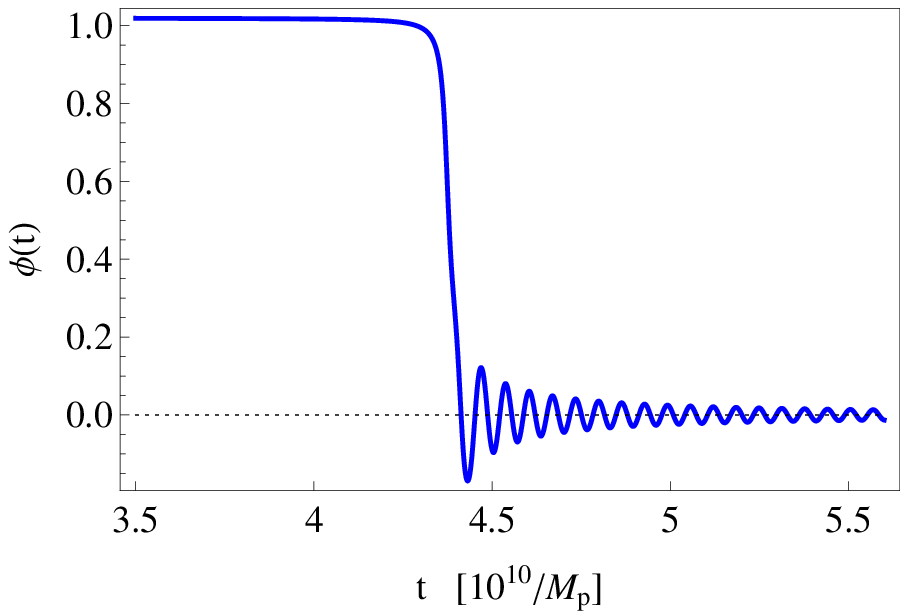}
          \hspace{1.0cm} $\phi(0)=1.0$ for $f=0.1$, \\$\Lambda_1/\Lambda_3=7.3$, $\Lambda_2/\Lambda_3=1$.
        \end{center}
      \end{minipage}
      \begin{minipage}{0.5\hsize}
        \begin{center}
          \includegraphics[clip, width=6.5cm]{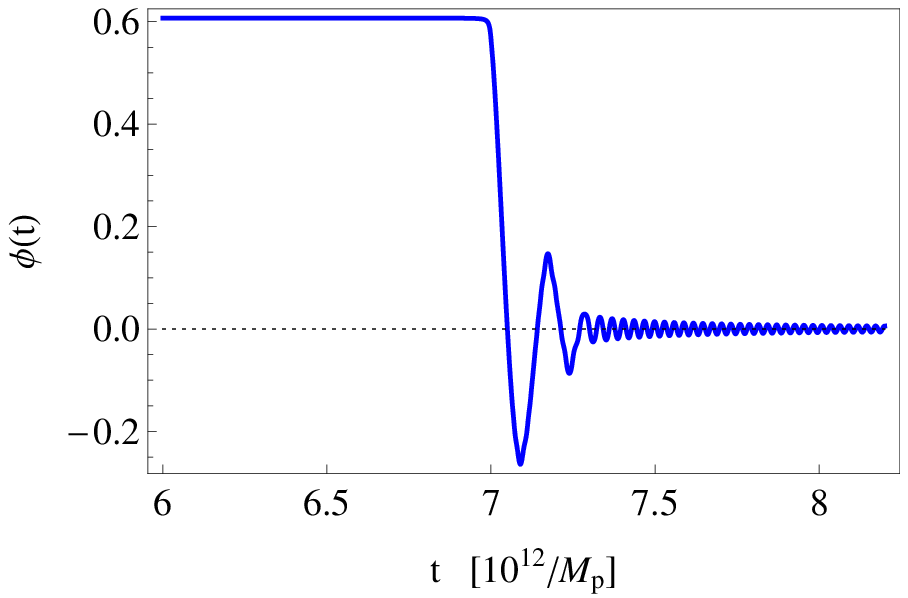}
          \hspace{1.0cm} $\phi(0)=0.6$ for $f=0.01$, $\Lambda_1/\Lambda_3=97$. $\Lambda_2/\Lambda_3=1$.
        \end{center}
      \end{minipage}
    \end{tabular}
    \caption{Inflaton trajectories. }
    \label{fig:numerical solution}
  \end{center}
\end{figure}

Fig. \ref{fig:numerical solution} shows examples of inflaton trajectories.
For the illustrative purpose, the initial values of the inflaton field are chosen such that a big enough e-folding number is realized at 
the second and tenth plateaus, respectively, for $f=0.1$ and $0.01$.
The inflaton rolls down through lower plateaus to finally reach the global minimum $\phi=0$.
The e-folding numbers, which are obtained from the other plateaus, are negligible for these examples. We concentrate on such parameter regions for concreteness where the total number of e-folds originates from a single plateau in the following discussions. We then aim to illustrate the characteristic features of our small field axion inflation scenarios which can be applicable for a wider range of the parameters.  

\begin{figure}[htbp]
  \begin{center}
    \begin{tabular}{c}
                    
      \begin{minipage}{0.5\hsize}
        \begin{center}
                    \includegraphics[clip, width=6.5cm]{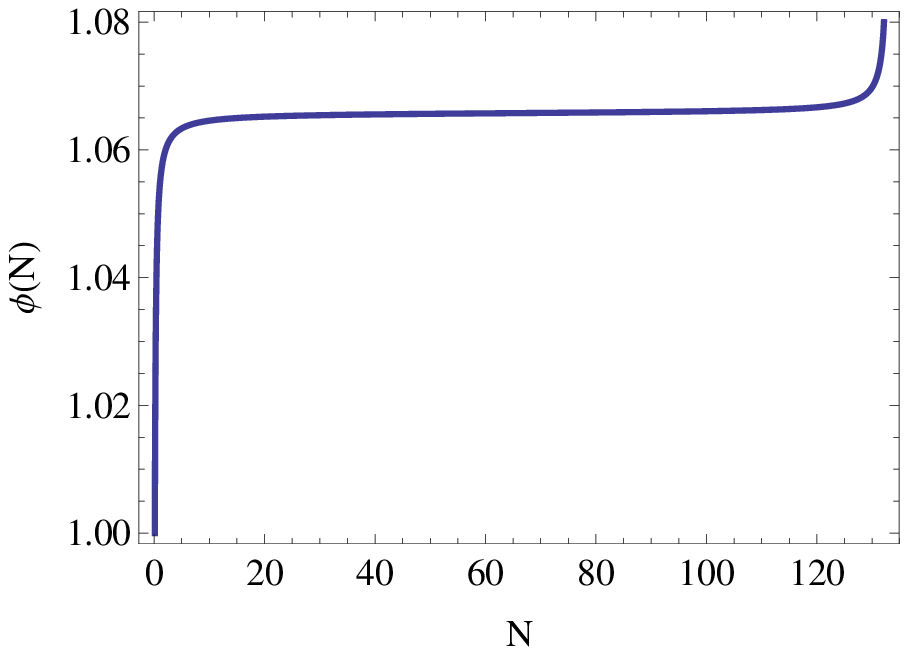}
                                 \hspace{1.0cm}  $ (N, \phi)$
        \end{center}
      \end{minipage}
              \begin{minipage}{0.5\hsize}
        \begin{center}
                    \includegraphics[clip, width=6.5cm]{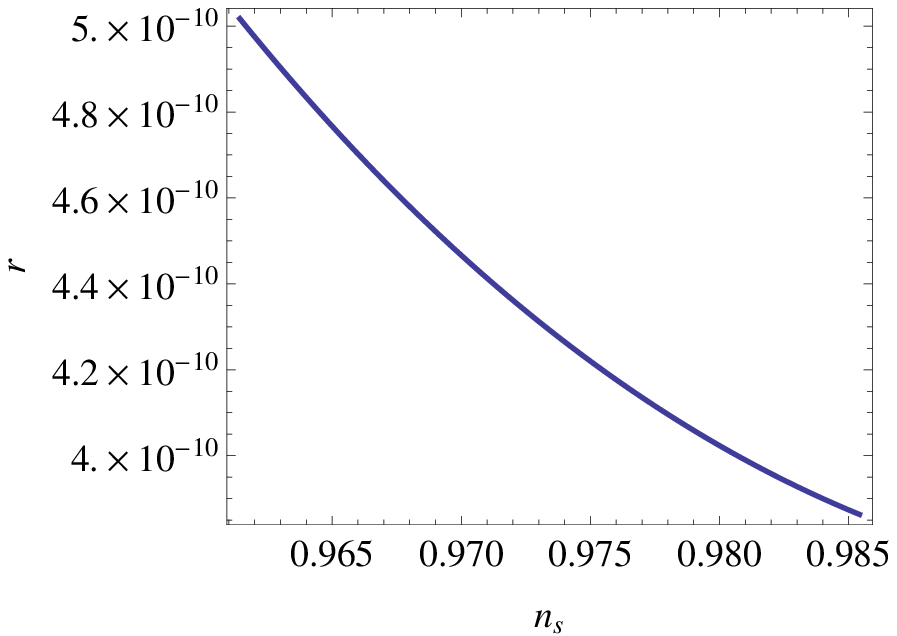}
                    \hspace{1.0cm}  $(n_s,r)$
        \end{center}
      \end{minipage}
    \end{tabular}
    \caption{ The inflaton amplitude as a function of the number of e-folds $\phi(N)$ (Left) and ($n_s$, $r$) for $N=[50,60]$ (Right) for $f=0.1$, $\Lambda_1/\Lambda_3=4.9$ and $\Lambda_2/\Lambda_3=0.25$ (corresponding to  $V^{1/4}_{\rm inf}(N=55)=6.1\times 10^{-5}$).}
    \label{fig:potential for f=0.1}
  \end{center}
\end{figure}

For $f=0.1$, Fig. \ref{fig:potential for f=0.1} shows how the inflaton field evolves as a function of the number of e-folds (counted from the end of inflation), and the corresponding tensor-to-scalar ratio $r$ and $n_s$ are shown. In Fig. \ref{fig:potential for f=0.1}, we consider the scenario where a sufficient number of e-folds are induced while the inflaton axion rolls over the second lowest plateau in the potential shown in Fig. \ref{fig:Vsmallf}. 
As reference values to indicate the energy scale of inflation, the Hubble parameter and the potential energy at $N=55$ in this example are $H_{\rm inf}(N=55)=2.2\times 10^{-9}$ and $\ V^{1/4}_{\rm inf}(N=55)=6.1\times 10^{-5}$.
The inflation on another plateau also can lead to a similar result, so that it can induce an enough number of e-folds from a single plateau with a small tensor-to-scalar ratio.

\begin{figure}[htbp]
  \begin{center}
    \begin{tabular}{c}
      \begin{minipage}{0.5\hsize}
        \begin{center}
                    \includegraphics[clip, width=6.5cm]{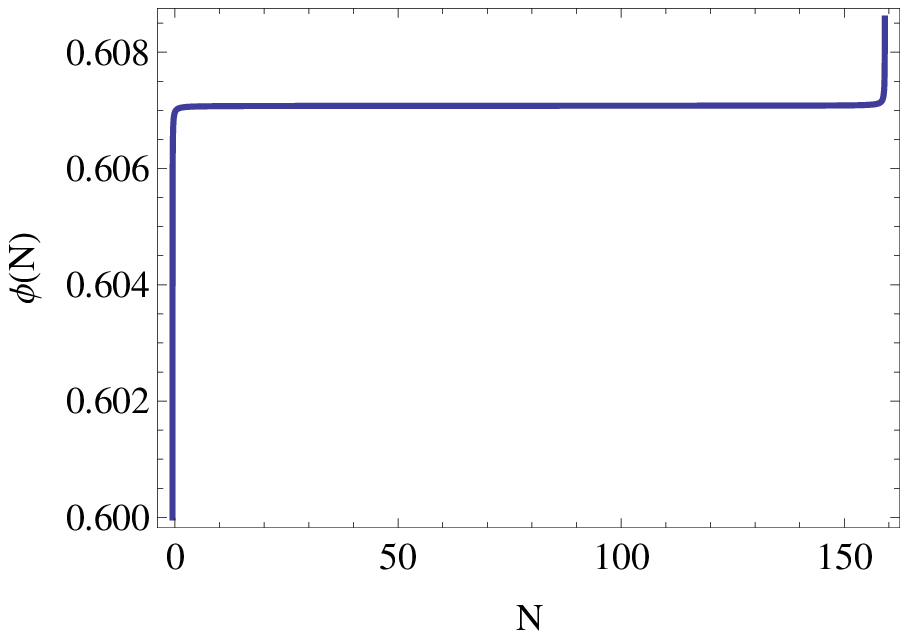}
             \hspace{1.0cm}  $ (N, \phi)$
        \end{center}
      \end{minipage}
              \begin{minipage}{0.5\hsize}
        \begin{center}
                    \includegraphics[clip, width=6.5cm]{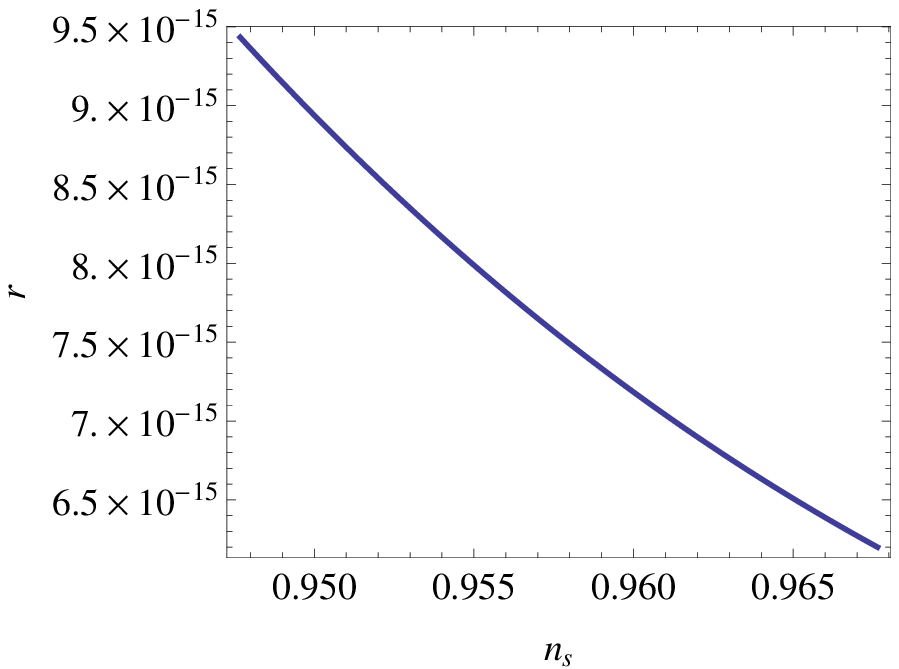}
                    \hspace{1.0cm}  $(n_s,r)$
        \end{center}
      \end{minipage}
    \end{tabular}
    \caption{The inflaton amplitude as a function of the number of e-folds $\phi(N)$ (Left) and ($n_s$, $r$) for $N=[50,60]$ (Right) for $f=0.01$, $\Lambda_1/\Lambda_3=97$, and $\Lambda_2/\Lambda_3=1$ (corresponding to $V^{1/4}_{\rm inf}(N=55)=4.0 \times 10^{-6}$). }
    \label{fig:potential for f=0.01}
  \end{center}
\end{figure}

The same story applies for a smaller $f=0.01$ as shown in Fig. \ref{fig:potential for f=0.01} (the scenario where a sufficient number of e-folds are induced on the tenth lowest plateau in Fig. \ref{fig:Vsmallf}) corresponding to $V^{1/4}_{\rm inf}(N=55)=4.0 \times 10^{-6}$ and $H_{\rm inf}(N=55)=9.0\times 10^{-12}$. 

\begin{figure}[htb]
  \begin{center}
    \begin{tabular}{c}
              \begin{minipage}{0.5\hsize}
        \begin{center}
                    \includegraphics[clip, width=6.5cm]{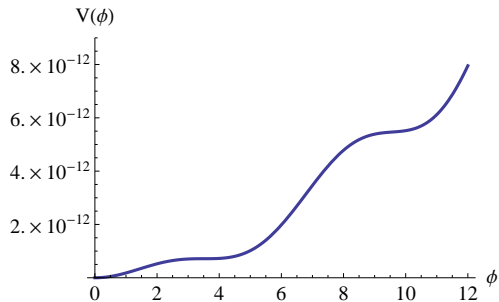}
        \end{center}
      \end{minipage}
    \end{tabular}
    \caption{The axion inflation potential with $f=1.0$ for the small field inflation (the field excursion $\Delta \phi<1$ during the inflation).}
    \label{fig:Vf-1.0-small}
  \end{center}
\end{figure}

\begin{figure}[htbp]
  \begin{center}
    \begin{tabular}{c}
                    
      \begin{minipage}{0.5\hsize}
        \begin{center}
                    \includegraphics[clip, width=6.5cm]{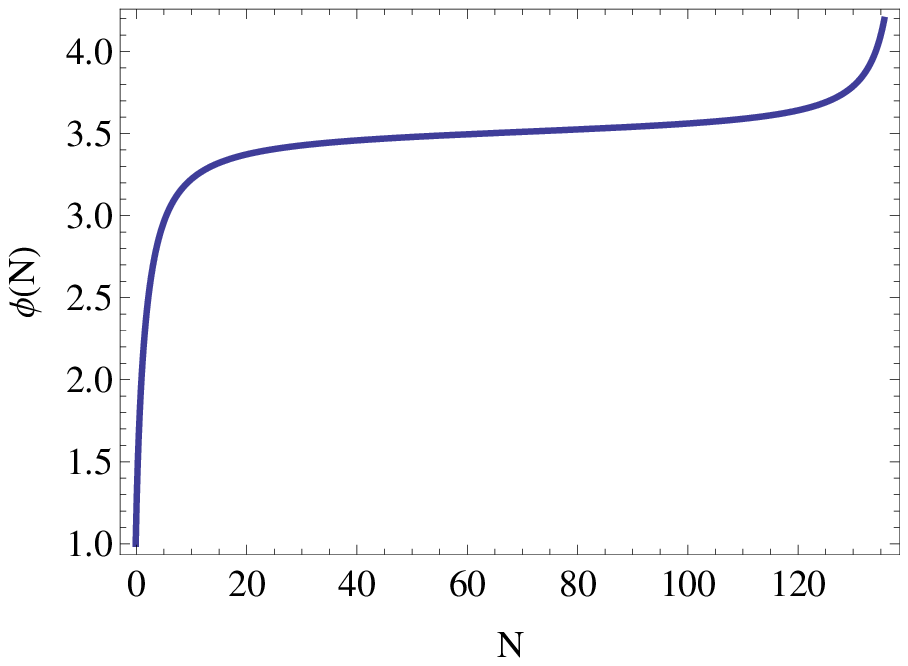}
             \hspace{1.0cm}  $ N\ {\rm vs}\ \phi(N)$
        \end{center}
      \end{minipage}
              \begin{minipage}{0.5\hsize}
        \begin{center}
                    \includegraphics[clip, width=6.5cm]{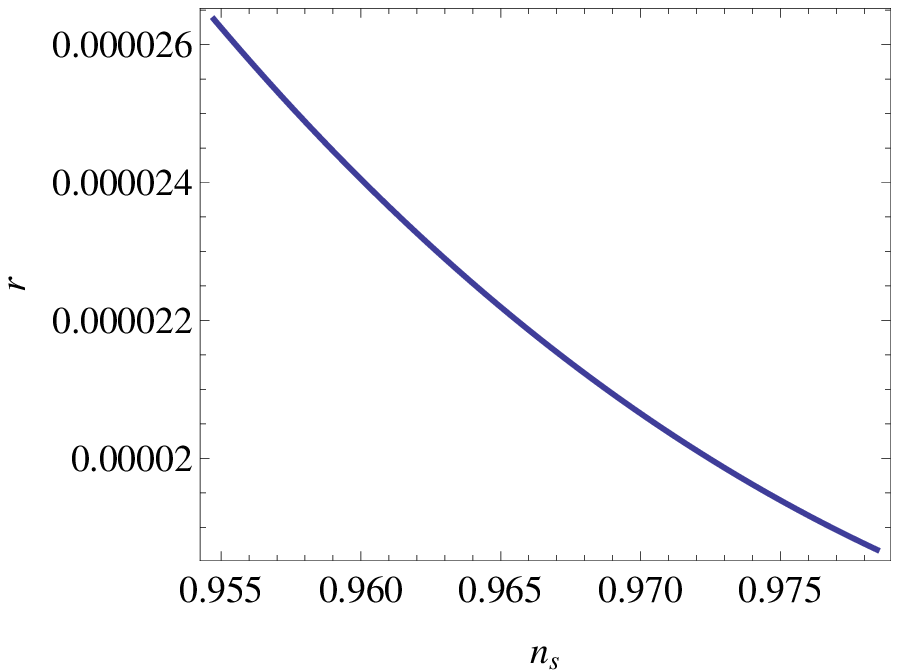}
                    \hspace{1.0cm}  $(n_s,r)$
        \end{center}
      \end{minipage}
    \end{tabular}
    \caption{(The inflaton amplitude as a function of the number of e-folds $\phi(N)$ (Left) and ($n_s$, $r$) for $N=[50,60]$ (Right) for $f=1.0$, $\Lambda_1/\Lambda_3=1.0$ and  $\Lambda_2/\Lambda_3=1.9$ (corresponding to $V^{1/4}_{\rm inf}(N=55)=9.0 \times 10^{-4})$.}
    \label{fig:potential for f=1.0}
  \end{center}
\end{figure}

For completeness, we also show the potential for $f=1.0$ in Fig. \ref{fig:Vf-1.0-small} and the evolution of $\phi$ along with $(n_s, r)$ in Fig. \ref{fig:potential for f=1.0} which corresponds to $V^{1/4}_{\rm inf}(N=55)=9.0 \times 10^{-4}$ and $H_{\rm inf}(N=55)=4.7\times 10^{-7}$. The inflaton field excursion during the inflation is sub-Planckian $\Delta \phi<1$ (we hence call it the small field inflation), even though the amplitude itself can be larger than the Planck scale.

The above numerical analysis demonstrates that our axion potential with a sub-Planckian axion decay constant 
as well as $f=1$ can lead to the inflation with a sub-Planckian inflaton field excursion. 
One notable feature compared with the conventional axion inflation scenarios with the Planckian $f$ and inflaton amplitude is a small tensor to scalar ratio $r\ll 1$. 
As discussed by Eq. (\ref{eq:r-approx}), $r$ is suppressed as the fourth power of $f$. A rough estimation Eq. (\ref{eq:r-approx}) fits with our numerical results by taking $\eta \sim 10^{-2}$ as mentioned above, and we estimate the typical parameter values of our axion inflation scenarios as 
\begin{equation}
r \sim 10^{-6}\times f^4, \qquad V^{1/4}_{\rm inf} \sim 5 \times 10^{-4} \times f,\qquad H_{\rm inf} \sim 10^{-7} \times f^2, 
\qquad \Lambda_3 \sim 6 \times 10^{-14} \times f^5, 
\label{eq:typical}
\end{equation}
because of $V_{\rm inf} \sim  \Lambda_3 /f$.
The energy scale of our axion inflation scenarios can be quite low compared with the conventional axion inflation with the Planckian decay constant, and we expect the consequent low reheating temperature as discussed in the next section.   

\begin{figure}[htb]
  \begin{center}
    \begin{tabular}{c}
                    
      \begin{minipage}{0.5\hsize}
        \begin{center}
                    \includegraphics[clip, width=6.5cm]{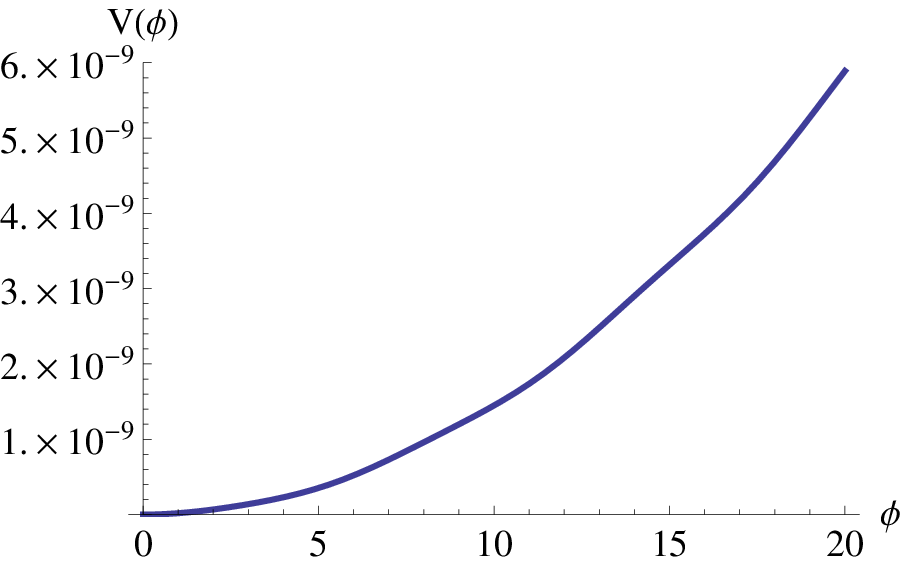}
                    \hspace{1.0cm}  $f=1$
        \end{center}
      \end{minipage}
              \begin{minipage}{0.5\hsize}
        \begin{center}
                    \includegraphics[clip, width=6.5cm]{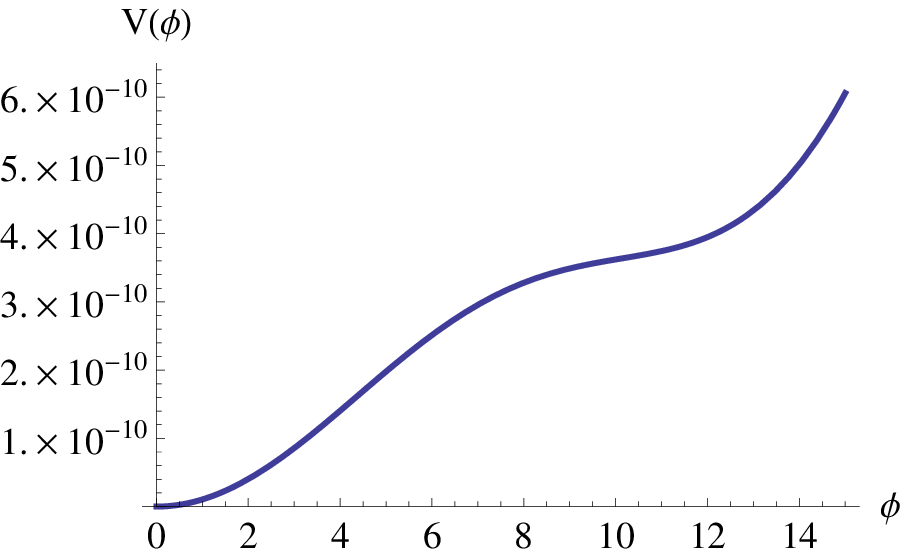}
             \hspace{1.0cm}  $f=3$
        \end{center}
      \end{minipage}
    \end{tabular}
    \caption{The axion inflation potential with a large axion decay constant for the large field inflation.}
    \label{fig:Vbigf}
  \end{center}
\end{figure}

Before concluding this section focusing a small $f$, let us briefly discuss the scenarios for a larger $f \gtrsim 1 $ commonly discussed in the literature for comparison. For a Planckian value of the axion decay constant, the large field inflation can be induced. The typical potentials are shown in Fig. \ref{fig:Vbigf} for $f=1$ and $f=3$. Compared with our axion potential with a sub-Planckian $f$, the tensor-to-scalar ratio $r$, along with the other parameters, can become large. For instance, with $f=3.0$ for concreteness, the first term $\Lambda_1 \phi^2$ can become dominant 
in both the potential (\ref{eq:original axion potential}) and the first derivative 
$V_{\phi}$ when $\phi \gg 1$ and $\Lambda_1 \sim \Lambda_2 \sim \Lambda_3$. The tensor-to-scalar-ratio ratio $r$ can be estimated as $r \sim 10/\phi^2$, e.g. $r\sim 0.1$ for $\phi \sim 10$. The representative examples for a Planckian $f$ are given, for illustration, in Fig. \ref{fig:Vbigf} and Table \ref{fig:bigf} showing the observables including the inflaton potential energy scale $V_{\rm inf}$ at the horizon exit $N=55$.

\begin{table}[htb] 
  \begin{tabular}{|c||c|c|c|c|c|c|} \hline
    & $ N$ & $n_s$  &  $r$ & $V^{1/4}_{\rm inf}$ & $\Lambda_1/\Lambda_3$ &  $\Lambda_2/\Lambda_3$ \\ \hline
        $f=1$ &  $ 55 $ & $0.95$  &  $0.13$ & $8.0 \times 10^{-3}$ &  $5.0$ & $1.0$  \\ \hline
        $f=3$ &  $ 55 $ & $0.97$  &  $0.011$ & $4.3 \times 10^{-3}$ &  $1.0$ & $4.9$  \\ \hline 
  \end{tabular}
    \caption{The typical parameters for $f=1,3$ for the large field inflation.}
    \label{fig:bigf}
\end{table}


\section{Phenomenology after inflation}


\begin{table}[h] 
  \begin{tabular}{|c|c||c|} \hline
     $f$  &   $m_{\phi}^2$ &$T_{\rm reh}$  \\ \hline
      $3.0$ & $1.0 \times 10^{-11}$ &  4.3  {\rm PeV} \\ \hline      
      $1.0$ & $1.9 \times 10^{-13}$ &  220 {\rm TeV} \\ \hline 
      $0.1$ & $1.2 \times 10^{-16}$ &  860 {\rm GeV} \\ \hline 
      $0.01$ & $3.4 \times 10^{-20}$ & 1.9 {\rm GeV} \\ \hline 
  \end{tabular}
    \caption{Typical reheating temperature for the cases $f=3.0,\ 1.0,\ 0.1,\ 0.01$ with $c=1$.}
    \label{fig:typical reheating temperature}
\end{table}

We now discuss the phenomenology after the inflation including the reheating temperature and the dark matter abundance in our small field axion scenarios.
The inflaton field is the axionic part of the complex structure modulus, and, in type IIB superstring theory, the complex structure moduli appear in one-loop corrections on gauge kinetic functions \cite{Lust:2003ky,Blumenhagen:2006ci}. The modulus thus couples with the gauge bosons through one-loop effects,
\begin{equation}
-\frac{1}{4g^2_a} F^a_{\mu \nu}F^{a \mu \nu} - \frac14 \frac{\Delta (\Phi)}{16\pi^2} F^a_{\mu \nu}F^{a \mu \nu},
\end{equation}
where $a=1,2,3$ correspond to the gauge groups of the standard model, 
$U(1)_Y$, $SU(2)$ and $SU(3)$, respectively, and $\Delta (\Phi)$ is a function of $\Phi$.
Through these couplings, the inflation decays into the gauge bosons $g^{(a)}$, 
and its decay width is estimated as \cite{Kobayashi:2015aaa}
\begin{eqnarray}
\Gamma_\phi & =&  \sum_{a=1}^3 \Gamma(\phi \rightarrow g^{(a)} + g^{(a)}) \nonumber \\
& =& \sum_{a=1}^3 \frac{N_G^a}{128 \pi} \left( \frac{\partial_\phi(\Delta(\Phi)) g_a^2}{16\pi^2 d}  \right)^2 
\frac{m_\phi^3}{M_{\rm p}^2} \\
& \simeq & 5.8 \times 10^{-5} c^2 \left( \frac{m_\phi}{10^{13}{\rm GeV}} \right)^3 {\rm GeV}, \nonumber
\end{eqnarray}
where $\sum_{a=1}^3 N_G^a = 12$, $d = {\cal O}(\sqrt{K_{\Phi \bar \Phi}}) ={\cal O}(1)$, $g_a^2 \simeq 0.53$, 
and  for concreteness, we assumed the form $\Delta(\Phi) = c\,\Phi$.
When such a decay into the gauge bosons is the dominant decay channel, the reheating temperature can be estimated as
\begin{eqnarray}
T_{\rm reh} = \left( \frac{\pi^2g_\ast}{90}\right)^{-1/4} \sqrt{\Gamma_{\phi}M_{\rm p}} \simeq 6.4 \times 10^{6}  c \left( \frac{m_{\phi}}{10^{13}\ {\rm GeV}}\right)^{3/2}{\rm GeV},
\label{eq:reheating temperature} 
\end{eqnarray}
where the effective degrees of freedom $g_*=106.75$.
Table \ref{fig:typical reheating temperature} lists the reheating temperature along with the inflaton mass for the concrete examples of $f=3.0,\ 1.0,\ 0.1,\ 0.01$ illustrated in the last section. A smaller $f$ corresponds to a smaller inflation energy scale, which hence leads to a smaller $T_{\rm reh}$. The order of magnitude for the inflaton mass can be estimated as follows.
For $f \ll 1$ with the relation  (\ref{eq:fine-tuning}), the dominant term of second derivative, $V_{\phi \phi}$, at $\phi =0$ is 
evaluated by $V_{\phi \phi} \sim \Lambda_3/f^2$, i.e. $m_\phi^2 \sim \Lambda_3 /f^2$.
Then, using Eq.~(\ref {eq:typical}), 
we can estimate the inflation mass by 
\begin{equation}
m_\phi^2 \sim 5 \times 10^{-14}\times f^3.
\label{eq:inflaton-mass}
\end{equation}

The complex structure moduli may appear in Yukawa couplings and higher dimensional couplings of matter fields 
within the framework of type IIB superstring theory (see for concrete computations, e.g. Ref.~\cite{Cremades:2004wa}).
The inflaton hence can also decay into the matter fields, and, when such a decay channel dominates, the reheating temperature can be estimated as \cite{Kobayashi:2015aaa}
\begin{eqnarray}
T_{\rm reh} 
 & \simeq & 8.8 \times 10^7 (\partial_\Phi Y_{ijk}) \left( \frac{m_\phi}{10^{13}{\rm GeV}} \right)^{3/2}  {\rm GeV},
\end{eqnarray}
where $\partial_\Phi Y_{ijk}$ denotes the first derivative of moduli-dependent Yukawa couplings $Y_{ijk}$.
$T_{\rm reh}$ estimated assuming the dominant decay via the Yukawa couplings is hence comparable or smaller than that estimated assuming the dominant decay into the gauge bosons.

Our models hence lead to the low reheating temperature (as low as GeV).
Such a low reheating temperature has important effects on the thermal history following the inflation. Dark matter relic abundance for instance could be affected significantly. 
For example, if the reheating temperature is smaller than the freeze-out temperature of dark matter, $T_{\rm reh} < T_{\rm f}$, the dark matter yield can be estimated by considering the non-thermal abundance from the inflaton decay  
\begin{align}
  \frac{n_{\rm dm}}{s}  \simeq \frac{n_{\rm inf}}{s}{\rm Br}_{\rm dm}
  \simeq\frac{\rho}{m_\phi s}\simeq 
\frac{3T_{\rm reh}}{4m_{\phi}} {\rm Br}_{\rm dm}\simeq 1.5\times 10^{-12}\left(\frac{c}{10}\right) 
\left( \frac{m_{\phi}}{10^{8}\ {\rm GeV}}\right)^{1/2}\left(\frac{{\rm Br}_{\rm dm}}{10^{-4}}\right),
 \label{eq:dmyield}
\end{align}
where $n_{\rm dm} (n_{\rm inf})$ is the number density of dark matter (inflaton), 
$s$ is the entropy density of the 
Universe, and 
${\rm Br}_{\rm dm}$ is the inflaton decay branching ratio to 
dark matter. 
The current dark matter abundance reads
\begin{align}
\Omega_{\rm dm}h^2 \simeq m_{\rm dm}\frac{n_{\rm dm}}{s} 
\frac{s_0}{\rho_{\rm cr}} 
\simeq 0.04\left(\frac{c}{10}\right)
\left(\frac{m_{\rm dm}}{100\,{\rm GeV}}\right) 
\left( \frac{m_{\phi}}{10^{8}\ {\rm GeV}}\right)^{1/2}\left(\frac{{\rm Br}_{\rm dm}}{10^{-4}}\right),
\end{align}
where $h$ denotes the dimensionless Hubble parameter and 
the ratio of critical density to the current entropy densities of the 
Universe is given by $\rho_{\rm cr}/s_0\simeq 3.6h^2\times 10^{-9}$. 

Our low energy scale axion inflation scenarios hence can be distinguished from the conventional large field axion inflation scenarios with a high reheating temperature $T_{\rm reh} > T_{\rm f}$ where the dark matter abundance can be estimated as the thermal relic abundance.

Another notable feature in our axion inflation scenarios with a small decay constant is the suppressed thermal production of the unwanted relics such as the gravitinos due to the low reheating temperature \cite{Kawasaki:2004yh}.
In general, supersymmetric models have the gravitino problem, 
and the low-energy effective field theory derived from superstring theory has the moduli problem. The non-thermally produced gravitinos from the moduli decay could be still a problem, and a light moduli, which does not contribute to 
supersymmetry breaking, can help in diluting the relic abundance of unwanted particles \cite{Akita:2016usy}.
The baryogenesis at a low temperature can be also a concern, and the low-energy scale Affleck-Dine mechanism can be a possibility in our scenarios to realize the desired baryon asymmetry of the Universe \cite{sky,jk}.




In addition to the inflaton axion we have been discussing so far, there can be other axion fields sourcing the isocurvature perturbations which give the tight bounds on the inflation parameters. For example, the isocurvature perturbations due to the QCD axion requires
\begin{equation}
H_{\rm inf} < 0.87 \times 10^{7} {\rm GeV} \left( \frac{f_a}{10^{11} {\rm GeV}} \right)^{0.408},
\end{equation}
where $f_a$ is the QCD axion decay constant (different from $f$), to be consistent with
the present observations \cite{Planck:2013jfk}.
Such a low scale inflation can be realized in our model with a sub-Planckian decay constant $f$.
For instance,  the models with $f=0.1$ and $0.01$ can lead to $H_{\rm inf} \sim 10^{9}$ GeV and $10^{7}$ GeV, respectively, 
while the model with $f=1.0$  leads to $H_{\rm inf} \sim 10^{12} $ GeV. It would be interesting to increase $f_a$, although there is an upper bound $f_a \lesssim 10^{12}$ to avoid the over-abundant axion while its precise upper bounds depend on the model details such as the initial displacement angles and the possible entropy dilution \cite{Akita:2016usy,Kawasaki:2004rx,Hattori:2015xla}. 

We so far limited our discussions to the case $ f \gtrsim 0.01$ as expected in the framework of type IIB superstring theory \cite{Kobayashi:2015aaa}. We could in principle study an even lower $f$, and compute the reheating temperature with Eqs.(\ref{eq:inflaton-mass}) and 
(\ref{eq:reheating temperature}).
However,  lower $f$ can, depending on $c$, lead to the reheating temperature of order MeV or below, and $ f \sim {\cal O}(0.01)$ would be the lower parameter range of our interest for the successful Big-Bang nucleosynthesis (BBN).


\section{Conclusion}

We have studied the axion inflation model proposed recently within the 
framework of type IIB superstring theory with a particular emphasis on the sub-Planckian axion decay constant, $0.01 \lesssim f \lesssim 1.0$. The axion potential with such a sub-Planckian decay constant possesses many flat plateaus and the small field inflation can be realized with a sufficient number of e-folds.

A notable feature of our scenario with a small decay constant $f$ is the low inflation energy scale $V_{inf} \propto f^4$ (Eq.~(\ref{eq:typical})). The implications of the consequent low reheating temperature in our string axion inflation scenarios were discussed including the dark matter abundance, gravitino/moduli problem and the isocurvature fluctuations of the QCD axion. More detailed studies  would be of great interest where we combine concrete mechanism for the moduli stabilization/uplifting, fix the mass scale of light moduli, choose a candidate for dark matter, and embed the QCD axion in superstring theory. We leave such detailed studies through the concrete models and their generalization for our future work.

We have studied one concrete potential which is derived from superstring theory. The shift symmetry of axion is violated by quantum effects inducing the axion potential. Such an axion potential consists of the mixture of polynomial functions and sinusoidal functions with 
the periodicity $\phi \sim \phi + 2\pi/f$, represented as $V(\phi^m, \cos(\phi/f), \sin(\phi/f))$.
For a small decay constant $f\ll 1$, such a potential can have many bumps and plateaus with the size of the flat regime $f/(2 \pi)$, and the small field inflation can be realized on one of the plateaus.

We expect our concrete examples discussed in our paper can capture the generic features for a wider class of axion inflation consisting of the sinusoidal and polynomial terms with a sub-Planckian axion decay constant.
For instance, let us assume that the sinusoidal parts are dominant in some derivatives of the potential. We then would find  $V^{(n+1)} \sim V^{(n)}/f$ with $n \geq n_{0}$ for a certain value $n_{0}$, where $V^{(n)}$ denotes the $n$-th derivative ($V^{(n+1)} \sim V^{(n)}/f$ can well happen for a higher derivative of the potential including the sinusoidal terms because a polynomial term vanishes at a sufficiently large $n$).
Analogous to Eq.~(\ref{eq:eta-f}), we can then make a similar Ansatz, $\eta \sim \Delta \phi f^{-p}$. Here, $p$ would depend on the form of the potential, e.g. $n_0$, while $p=2$ in our model presented in this paper.
This would lead to $r \sim 10^{-6} \times f^{2p}$ when the tensor-to-scalar ratio is small $r < {\cal O}(10^{-2})$ and we can estimate $2 \eta \approx n_s -1 \approx 0.03$.
In such a model analogous to ours discussed in this paper, the inflation energy scale and Hubble parameter could have the power law dependence on $f$ and hence become rapidly small as $f$ becomes small. As a consequence, the reheating temperature would become small too although its precise value depends on the detailed reheating processes such as couplings between the inflaton and light modes.
We would also be able to put the tight lower bound on $f$ from the BBN so that $T_{\rm reh} > {\cal O}(1)$ MeV.
Confirming such a generalization of our study is beyond the scope of current work, and we plan to present the analysis extending our studies here for a wider class of axion inflation models which can be explicitly derived from superstring theory in our future work.


\section*{Acknowledgments}
This work was in part supported by Institute for Basic Science (IBS-R018-D1) (KK), Grant-in-Aid for Scientific Research No.~25400252 (TK) and No.~26247042 (TK) from the Ministry of Education, Culture, Sports, Science and Technology (MEXT) in
Japan.

%





\begin{thebibliography}{99}



\bibitem{Freese:1990rb} 
  K.~Freese, J.~A.~Frieman and A.~V.~Olinto,
  Phys.\ Rev.\ Lett.\  {\bf 65}, 3233 (1990).

\bibitem{Svrcek:2006yi} 
  P.~Svrcek and E.~Witten,
  JHEP {\bf 0606}, 051 (2006)
  [hep-th/0605206].



\bibitem{Kim:2004rp}
  J.~E.~Kim, H.~P.~Nilles and M.~Peloso,
  JCAP {\bf 0501} (2005) 005
  [hep-ph/0409138].
  

\bibitem{Abe:2014pwa}
  H.~Abe, T.~Kobayashi and H.~Otsuka,
  PTEP {\bf 2015} 6,  063E02
  [arXiv:1409.8436 [hep-th]].
  



\bibitem{Abe:2014xja}
  H.~Abe, T.~Kobayashi and H.~Otsuka,
  JHEP {\bf 1504} (2015) 160
  [arXiv:1411.4768 [hep-th]].
  



\bibitem{Silverstein:2008sg} 
  E.~Silverstein and A.~Westphal,
  Phys.\ Rev.\ D {\bf 78}, 106003 (2008)
  [arXiv:0803.3085 [hep-th]];
%

  L.~McAllister, E.~Silverstein and A.~Westphal,
  Phys.\ Rev.\ D {\bf 82}, 046003 (2010)
  [arXiv:0808.0706 [hep-th]];

  R.~Flauger, L.~McAllister, E.~Pajer, A.~Westphal and G.~Xu,
  JCAP {\bf 1006}, 009 (2010)
  [arXiv:0907.2916 [hep-th]];

  X.~Dong, B.~Horn, E.~Silverstein and A.~Westphal,
  Phys.\ Rev.\ D {\bf 84}, 026011 (2011)
  [arXiv:1011.4521 [hep-th]].


\bibitem{Kobayashi:2014ooa} 
  T.~Kobayashi, O.~Seto and Y.~Yamaguchi,
  PTEP {\bf 2014}, no. 10, 103E01 (2014)
  [arXiv:1404.5518 [hep-ph]].

\bibitem{Higaki:2014sja} 
  T.~Higaki, T.~Kobayashi, O.~Seto and Y.~Yamaguchi,
  JCAP {\bf 1410}, no. 10, 025 (2014)
  [arXiv:1405.0775 [hep-ph]].





\bibitem{Lyth bound}
David H. Lyth, 
Phys. Rev. Lett. {\bf 78} 1861 (1997) [hep-ph/9606387].


 
\bibitem{Peloso:2015dsa} 
  M.~Peloso and C.~Unal,
  JCAP {\bf 1506}, no. 06, 040 (2015)
  [arXiv:1504.02784 [astro-ph.CO]].
 

\bibitem{Kobayashi:2016mzg} 
  T.~Kobayashi, D.~Nitta and Y.~Urakawa,
  arXiv:1604.02995 [hep-th].





\bibitem{Kobayashi:2015aaa}
  T.~Kobayashi, A.~Oikawa and H.~Otsuka,
  Phys.\ Rev.\ D {\bf 93} (2016) no.8,  083508
  [arXiv:1510.08768 [hep-ph]].
 
 

\bibitem{Hebecker:2015rya} 
  A.~Hebecker, P.~Mangat, F.~Rompineve and L.~T.~Witkowski,
  Phys.\ Lett.\ B {\bf 748}, 455 (2015)
  [arXiv:1503.07912 [hep-th]].
 
 
\bibitem{Parameswaran:2016qqq} 
  S.~Parameswaran, G.~Tasinato and I.~Zavala,
  JCAP {\bf 1604}, no. 04, 008 (2016)
  [arXiv:1602.02812 [astro-ph.CO]].
  
\bibitem{Bizet:2016paj}
  N.~C.~Bizet, O.~Loaiza-Brito and I.~Zavala,
  arXiv:1605.03974 [hep-th].





\bibitem{Planck deta}
Planck Collaboration: P. A. R. Ade {\it et al},
arXiv:1502.02114 [astro-ph. CO]. 
\bibitem{plbi} 
  P.~A.~R.~Ade {\it et al.} [BICEP2 and Planck Collaborations],
  Phys.\ Rev.\ Lett.\  {\bf 114}, 101301 (2015)
  [arXiv:1502.00612 [astro-ph.CO]].



  

\bibitem{Giddings:2001yu} 
  S.~B.~Giddings, S.~Kachru and J.~Polchinski,
  Phys.\ Rev.\ D {\bf 66}, 106006 (2002)
  [hep-th/0105097].

\bibitem{Gukov:1999ya} 
  S.~Gukov, C.~Vafa and E.~Witten,
  Nucl.\ Phys.\ B {\bf 584}, 69 (2000)
  Erratum: [Nucl.\ Phys.\ B {\bf 608}, 477 (2001)]
  [hep-th/9906070].


\bibitem{Kachru:2003aw} 
  S.~Kachru, R.~Kallosh, A.~D.~Linde and S.~P.~Trivedi,
  Phys.\ Rev.\ D {\bf 68}, 046005 (2003)
  [hep-th/0301240].


\bibitem{Dudas:2006gr} 
  E.~Dudas, C.~Papineau and S.~Pokorski,
  JHEP {\bf 0702}, 028 (2007)
  [hep-th/0610297];

  H.~Abe, T.~Higaki, T.~Kobayashi and Y.~Omura,
  Phys.\ Rev.\ D {\bf 75}, 025019 (2007)
  [hep-th/0611024];

  R.~Kallosh and A.~D.~Linde,
  JHEP {\bf 0702}, 002 (2007)
  [hep-th/0611183];

  H.~Abe, T.~Higaki and T.~Kobayashi,
  Phys.\ Rev.\ D {\bf 76}, 105003 (2007)
  [arXiv:0707.2671 [hep-th]].



\bibitem{Blumenhagen:2016bfp}
  R.~Blumenhagen, D.~Herschmann and F.~Wolf,
  arXiv:1605.06299 [hep-th].




\bibitem{Lust:2003ky} 
  D.~Lust and S.~Stieberger,
  Fortsch.\ Phys.\  {\bf 55}, 427 (2007)
  [hep-th/0302221].



\bibitem{Blumenhagen:2006ci} 
  R.~Blumenhagen, B.~Kors, D.~Lust and S.~Stieberger,
  Phys.\ Rept.\  {\bf 445}, 1 (2007)
  [hep-th/0610327].

\bibitem{Cremades:2004wa} 
  D.~Cremades, L.~E.~Ibanez and F.~Marchesano,
  JHEP {\bf 0405}, 079 (2004)
  [hep-th/0404229];
%

  H.~Abe, K.~S.~Choi, T.~Kobayashi and H.~Ohki,
  JHEP {\bf 0906}, 080 (2009)
  [arXiv:0903.3800 [hep-th]].


\bibitem{Kawasaki:2004yh} 
  M.~Kawasaki, K.~Kohri and T.~Moroi,
  Phys.\ Lett.\ B {\bf 625}, 7 (2005)
  [astro-ph/0402490];
%
%
  Phys.\ Rev.\ D {\bf 71}, 083502 (2005)
  [astro-ph/0408426].


\bibitem{Akita:2016usy} 
  K.~Akita, T.~Kobayashi, A.~Oikawa and H.~Otsuka,
  JHEP {\bf 1605}, 178 (2016)
  [arXiv:1603.08399 [hep-ph]].


\bibitem{sky} 
  E.~D.~Stewart, M.~Kawasaki and T.~Yanagida,
  Phys.\ Rev.\ D {\bf 54}, 6032 (1996)
  [hep-ph/9603324].
  

\bibitem{jk} 
  D.~h.~Jeong, K.~Kadota, W.~I.~Park and E.~D.~Stewart,
  JHEP {\bf 0411}, 046 (2004)
  [hep-ph/0406136].


\bibitem{Planck:2013jfk} 
  P.~A.~R.~Ade {\it et al.} [Planck Collaboration],
  Astron.\ Astrophys.\  {\bf 571}, A22 (2014)
  [arXiv:1303.5082 [astro-ph.CO]].


\bibitem{Kawasaki:2004rx} 
  M.~Kawasaki and F.~Takahashi,
  Phys.\ Lett.\ B {\bf 618}, 1 (2005)
  [hep-ph/0410158].

\bibitem{Hattori:2015xla} 
  H.~Hattori, T.~Kobayashi, N.~Omoto and O.~Seto,
  Phys.\ Rev.\ D {\bf 92}, no. 10, 103518 (2015)
  [arXiv:1510.03595 [hep-ph]].





\end{thebibliography}
\end{document}